\documentstyle[epsf]{mn}
\title{On the Electric Field Screening by Electron-Positron Pairs
in the Pulsar Magnetosphere II}
\author[S. Shibata et al.]
       {S. Shibata$^1$, J. Miyazaki$^2$, and F. Takahara$^3$ \\
        $^1$Department of Physics, Yamagata University, 
        1-4-12 Kojirakawa, Yamagata 990, Japan \\
        $^2$Department of Physics, Tokyo Metropolitan University, 
        Minamiohsawa 1-1, Hachioji, Tokyo 192-03, Japan \\
        $^3$Department of Earth and Space Science, Graduate School of
	Science, Osaka University, Toyonaka, Osaka 560-0043, Japan
	}

\date{Accepted --- ; in draft form 2001 September 14; revised on 27 May 2002}

\newcommand{\eq}[1]{\label{#1} }                             
\begin{document}

\maketitle

\begin{abstract}
We study on the self-consistency of the pulsar polar cap model, i.e.,
the problem of whether the field-aligned electric field is screened by
electron-positron pairs that are injected beyond the pair
production front.
We solve the one-dimensional Poisson equation along a magnetic field line,
both analytically and numerically, for a given current density
incorporating effects of returning positrons, and we obtain the conditions for
the electric-field screening. 
The formula which we obtained gives the screening distance and the
return flux for given primary current density, field geometry and 
pair creation rate at the pair production front.

If the geometrical screening  is not possible,
for instance, on field lines with a super-Goldreich-Julian current,
then the electric field at the pair production
front is constrained 
to be fairly small in comparison with values expected typically
by the conventional polar cap models.
This is because
(1) positive space charge by pair polarization is limited to a small value,
and 
(2) returning of positrons leave pair electrons behind.
A previous belief that pair creation
with a pair density higher than the Goldreich-Julian density immediately
screens out the electric field is unjustified at least for
for a super Goldreich-Julian current density.
We suggest some possibilities to
resolve this difficulty.
\end{abstract}

\section{Introduction}

One of the promising models for the pulsar action is the polar cap model
where the field-aligned electric field accelerates charged particles 
up to TeV energies, and resultant curvature radiation generates 
an electromagnetic shower. 
The model can provide an environment for radio emission mechanism
with a particle beam in electron-positron pair plasmas and 
may also explain the gamma-ray pulsation.
The pairs will be a particle source of the pulsar wind.
The polar cap potential drop is a part of the electromotive force
produced by the rotating magnetic neutron star. 
The voltage is a few percent of the available voltage for young pulsars,
while it becomes some important fraction for older pulsars.
The localized potential drop is maintained by a pair of anode and 
cathode regions, the formation mechanism of which is the long-standing issue
of the polar cap accelerator.

Because any deviation of the space charge density from the Goldreich-Julian
(GJ) density $\rho_{\rm GJ}$
causes the field-aligned electric field
within the light cylinder, 
how the GJ charge density changes along a given magnetic field line plays
an essential role in formation of the field-aligned electric field.
Note here that the GJ density is determined by the magnetic field geometry:
$\rho_{\rm GJ} \approx - \Omega B_z / 2 \pi c$, where $\Omega$ is the 
angular velocity of the star and $B_z$ is the field strength along the
rotation axis.
For instance, the outer gap, another promising accelerator model,
is thought to appear around the null surface on which the GJ density
changes its sign. Due to the field geometry, 
the outer gap originally has a pair
of the anode and cathode regions without current and pair plasma.
The space-charge-limited flow for the polar cap model on field lines
curving toward the rotation axis also has a similar GJ density distribution
providing a pair of the anode and cathode regions
(actually it is regarded as a version of outer gap with 
external current [Hirotani \& Shibata 2001]).

Once pairs are created in the accelerator, since
the pair {\em number density} far exceeds 
the GJ value $|\rho_{\rm GJ}/e|$,
where $-e$ is the electronic charge,
dynamics of the created pairs makes crucial effects in
localizing the field-aligned electric field.

For the polar cap models, the space-charge-limited flow
is found to produce the field-aligned electric field
in some cases where the current density and
magnetic field geometry are assumed
(Fawley, Arons \& Scharlemann 1977, 
Scharlemann, Arons \& Fawley 1978).
Later on, effects of pair creation and general relativity are
taken into account
(Arons \& Scharlemann 1979, Muslimov \& Tsygan 1992).
In these models,
given a current density, the space charge density deviates from
the GJ charge density, and as a result the field-aligned electric
field develops. Since the work function of charged particles on
the stellar surface is sufficiently small, either electrons or ions can
freely escape from the surface, 
and therefore the electric field on the surface is
assumed to be zero. Based on this scheme, primary electrons are shown to
be accelerated to energies high enough to induce
an electromagnetic shower.

The current density on each magnetic lines of force will not be
determined by the local electrodynamics in the polar cap, but 
will be determined in a global manner.
Most of the rotation power of the pulsar is conveyed by the pulsar wind,
which is accelerated near and beyond the light cylinder. Although
the acceleration mechanism of the wind is still mystery, it is certain
that an energetically dominant process which determines the current distribution
takes place near and beyond the light cylinder.
When the wind is accelerated (or heated), the current runs across
the field lines ($\vec{E} \cdot \vec{j} >0$) as a simple result
of the Poynting theorem, provided that the ideal-MHD condition is
more or less valid. Therefore, some of the current coming up from the star
crosses the magnetic field lines corresponding to the wind acceleration and
returns back to the star.
The rest of the current reaches large distances (e.g., nebula shock), and 
the same amount with opposite direction comes back to the star.
In any case, the current closure is seriously controlled by 
the dynamics near and beyond the light cylinder.
The current determined in such a way runs through the polar cap.
Thus, the current density distribution across the  polar-cap magnetic flux 
should be linked to the outer magnetosphere:
the current density distribution 
is most likely to be determined regardless of the
field geometry of the inner magnetosphere. 
We, therefore, treat the current density as an adjustable free
parameter in our series of paper 
(Shibata 1991, 1995, 1997, Shibata, Miyazaki \& Takahara 1998):
the current density is a free parameter for the polar cap model, 
and it is adjusted when 
the local model is linked with a model of the outer magnetosphere.

For a space-charge-limited flow, it is shown by Shibata (1997)
that both toward and away geometries
are possible to generate a large field-aligned electric field
causing an electron-positron pair cascade
as long as the current density is reasonably large, comparable
to the GJ current density $\rho_{\rm GJ}c$. 
In his paper, the cases of super-GJ current densities are
examined as well as the sub-GJ cases.
Note that the GJ current density is simply 
the GJ current, $I_{\rm GJ} \approx \mu \Omega^2 /c$
divided by the polar cap area,
where $\mu$ is the magnetic moment of the star.
The GJ current is derived by an order-of-magnitude
argument, and the net current circulating in the magnetosphere
will not be far different from this value. 
However, if the current is 
not distributed more or less uniformly over the polar cap, 
but is focused or fragmented in the magnetic flux tubes
by analogy of planetary aurori, 
then the local current densities can be much higher than
the GJ current density. 
Furthermore, the current density will change in a wide rage
field lines by field line.

These points motivate us to examine various 
combinations of current density
and magnetic field geometry for the local accelerator.
Although deviation of the space charge is found to cause the accelerating
electric field, it is not clear how the electric field is screened
to complete the localized potential drop.
Only the known case in which a finite potential drop is formed without pairs 
is the geometrical screening, i.e.,
the case of toward geometry or general relativistic geometry
(Muslimov \& Tsygan 1992)
with a certain current density which is sub-GJ. 
For other cases, there remains in general an unscreened electric field
in a region where pairs are created.
Note that pair plasmas are copious and pair polarization may
become efficient.
In this paper, we derive the condition for the electric-field
screening immediately behind the pair creation front where unscreened
electric field is assumed to exist, in the presence of
returning pair.
The condition can be used for any current density (super- or sub-GJ)
and for both away and toward field line curvatures.
We will apply the condition to the cases for which geometrical effect
is not efficient, i.e., away curvature and super-GJ current density.
Another important application is the case where pairs are
created before the geometrical screening completes.


In the previous paper (Shibata, Miyazaki \& Takahara 1998,  Paper I), 
we demonstrated that the pair polarization does work when enough 
pairs are injected, where we assumed that pairs are created at a single point; 
the required multiplication factor is $\sim 10^3$ which is
the same order of the value inferred from observations of the Crab
pulsar and nebula. 
However, we suggested that the required multiplicity has to be
reached in a very small distance, and this will not be realized
in an actual pulsar magnetosphere.

We have made some simplifications in Paper I in order to make
the physics as clear as possible. They are,
1) pairs have the same initial Lorentz factor and outward momentum,
2) the pair creation occurs at a single point, and
3) no positrons return to the neutron star.
In this paper, we relax the above assumptions 2) and 3) to study more
realistic cases: electron-positron pairs are produced continuously
along field lines for a finite length, and
some particles may stop before reaching the screening point
and return to the neutron star surface.
As in Paper I, we assume that
no frictional force works between the various components.
Under these assumptions, we calculate the electric field structure of
the pair creation zone and determine the maximum electric field
which can be screened under a presence of returning particles.
Since the thickness of this zone is geometrically thin, we
adopt one-dimensional approximation in this paper.

In section 2, we formulate the problem, and in section 3 we solve the
one-dimensional Poisson equation analytically assuming that
the pair creation rate is constant as a function of the electrostatic
potential. In section 4, we solve the problem numerically
when the pair creation rate is spatially uniform
to confirm that the analytic results of section 3 are sufficiently 
accurate.
In section 5 we summarize our results 
and suggest a possible way out to resolve the screening problem.

\section{Model and Basic Equations}

Let us consider a steady model of a field-aligned accelerator which
has a finite potential drop with the electric field screened at the both ends.
For definite sign of charge, let us assume electrons are accelerated
outwards, i.e., the electric field points toward the star.
Gamma-rays  emitted by the electrons  convert into pairs
beyond a certain surface called
the pair production front (PPF);  beyond PPF pairs are assumed to 
be created continuously in space.

Pair polarization is an effect that screens out the electric
field. 
The pair positrons are decelerated soon after their birth,
while the pair electrons are accelerated, 
and as a result of continuity,
a positive space charge appears to reduce the electric field.
In this paper, we intend to evaluate the space charge produced
by pair polarization
in the presence of returning positrons.

Notations are the same as Paper I:
the strength of the magnetic field $\vec{B}$ is denoted by $B$, and
the component along the rotation axis is $B_{\rm z}$;
the non-corotational electric potential $\Phi$ gives the
field-aligned electric field by 
$E_\parallel = - {\vec B} \cdot \nabla \Phi /B$;
$v$ and $\rho$ are the velocity and density with 
the subscripts `1', `$+$' and `$-$' indicating
the primary particles, pair positrons and pair electrons,
respectively;
$\rho_{\rm GJ} \approx - (\Omega B_{\rm z})/2 \pi c$ 
is the GJ charge density.
We also use the following normalized quantities:
\begin{eqnarray}
\phi & = & \frac{e\Phi}{mc^2}, \nonumber \\
j&=&J/(-\frac{\Omega B}{2\pi}), \nonumber \\
j_0&=&c \; \rho_{\rm GJ} /(-\frac{\Omega B}{2\pi}) = B_{\rm z}/B , \nonumber \\
\bar{\rho} & = & \rho / ({\Omega B \over 2 \pi c} ) \nonumber \\
\beta&=&\frac{|v|}{c}, \quad \gamma = (1-\beta^2 )^{-1/2} \nonumber \\
x&=&\frac{s\omega_p}{c},\quad
\omega_p=\sqrt{\frac{4\pi e^2n_0}{m}}, \quad
n_0=\frac{\Omega B}{2\pi ec}, \nonumber
\end{eqnarray}
where $J= - e n_1 v_1$ is the current density of the primary electrons; 
$s$ is the coordinate along the magnetic field, which is normalized
by the relativistic Debye length.

The Poisson equation for the screening region may be
\begin{equation}
-\frac{d^2\phi}{dx^2}=\frac{dE}{dx}=
-\frac{j}{\beta_1}+j_0+\bar{\rho}_++\bar{\rho}_-,
\eq{Poisson1}
\end{equation}
where $E =-d \phi /dx$ is the normalized electric field along                   a given magnetic field,
$\beta_1$ is the normalized velocity of the primary particles.
In the right hand side of (\ref{Poisson1}),
the first term represents the negative space charge produced
by the primary electronic current stream ($-j/\beta_1 < 0$), and 
$j_0$ represents $- \rho_{\rm GJ}$, which is positive.
The difference, $-j/ \beta_1 +j_0$, is the space charge 
that produces the accelerating electric field.
The values $\bar{\rho}_\pm$ are
the normalized charge densities of the pair electrons and pair positrons,
which appear  beyond PPF.
Note that $j$ and $j_0$ take positive values in our sign convention.

In a steady acceleration region along given magnetic field lines,
`an effective space charge' which is represented by the right hand side
of (\ref{Poisson1}) becomes positive or negative at each side of
the accelerating region as shown in Figure~1.
The negative charge region is located near the stellar surface and
electrons are accelerated upwards.
In the space-charge-limited flow
(Fawley, Arons \& Scharlemann, 1977), 
the negative charge is
provided by non-relativistic electrons leaving the stellar surface;
$-j/\beta_1 + j_0 <0$ with $\beta_1 \ll 1$.
Additional negative charge is induced on the side wall of the flow tube.
As the electrons are accelerated, $\beta_1 \rightarrow 1$,
the effective space charge becomes positive above a certain height, 
$-j+j_0 >0$,
if $j$ is smaller than $j_0$.
Thus, the anode region is formed in this case.

Further development is to include the effect of magnetic field geometry:
on field lines curving toward the rotation axis,
$j_0 = B_{\rm z}/B$ ($<1$) increases along the field lines
while $j \propto J/B$ is constant, and
therefore even if $-j+j_0 <0$ near the star,
as the flow goes out, $-j+j_0$ can be positive above some distance
as far as $j<1$ (sub-GJ).
This toward curvature is efficient to screen the electric field
which is sufficiently strong to cause pair cascade.

One can choose a particular value of $j$ so that the electric field
is localized with the screened electric field at the both ends without
pairs. However,
as has been mentioned, the current density is determined in a global way.
It is therefore unlikely that such a particular value of the current density
is realized in general.
One can expect that PPF is formed at the place where the electric field
remains unscreened.

If $j>1$, the anode formation by the geometrical effect
does not work because $-j+j_0<0$.
The same situation takes place for any value of $j$ on field lines
curving away from the rotation axis
(Mestel 1981).
In these cases, the steady accelerator is possible to exist
only if an anode region is provided in some way for which
we study the dynamics of pairs in the PPF region.
%
\begin{figure}
\begin{center}
\mbox{\epsfysize=5.0cm \epsfbox{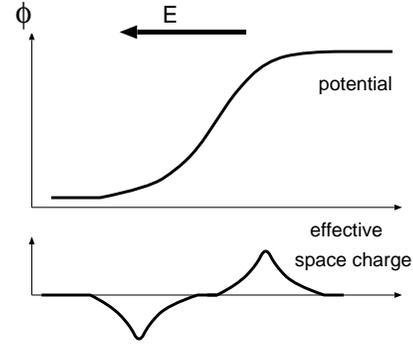}}
\end{center}
\caption{%
An accelerating region with a finite potential drop has
a pair of anode and cathode regions.
For a parallel rotator (not anti-parallel),
negative space charge appears in the side of the stellar surface,
and positive space charge is required for `screening' the electric
field at the other side of the potential drop. 
The positive screening charge can be provided by toward curvature
(on field lines curving toward the rotation axis) and by
a pair plasma. 
}
\end{figure}


Since screening by pairs occurs in a small scale such
as $\gamma mc^2/e|E_\parallel |$,
$j$ and $j_0$ are assumed to be constant in the screening region
around PPF.


\section{Condition for Screening}

The screening region is schematically shown in Figure~2.
The pair creation starts at PPF ($x=0$), 
and continues beyond ($x \geq 0$).
The non-dimensional electric field at $x=0$ is denoted by $- E_{\rm f}$, and
the potential at PPF is used for the reference, i.e.,
$\phi = 0$ at $x=0$.
Let us assume that the field-aligned electric field is 
screened out at a certain point defining the screening surface (SS).

\begin{figure}
\begin{center}
\mbox{\epsfysize=11cm \epsfbox{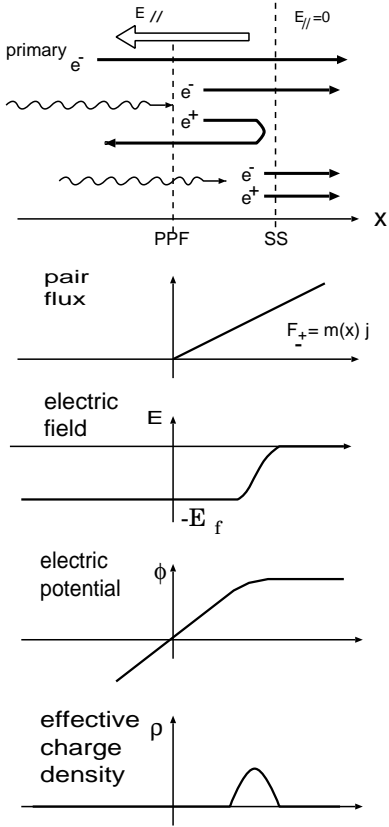}}
\end{center}
\caption{%
Schematic diagram of the screening region by pair polarization.
Electron-positron pair plasma is created at more or less
uniform rate beyond the pair production front (PPF).
The pair positrons are decelerated while the pair electrons are
accelerated. When the pair positrons become non-relativistic,
there appears positive space charge, which should screen out
the accelerating electric field.
Some of pair positrons will turn back to the star.
Pairs created near the screening surface (SS) go through
the region.
}
\end{figure}

Multiplying (\ref{Poisson1}) by $E$ and
integrating it from PPF to SS, we obtain the screening condition
\begin{equation}
\frac{1}{2}E_{\rm f}^2=\int^{\phi_{\rm s}}_0(\bar{\rho}_+
+\bar{\rho}_-)d\phi-(j-j_0)\phi_{\rm s},
\eq{Poisson2}
\end{equation}
where we have assumed for the primary particles $\beta_1=1$,
and the subscripts `s' indicates values at SS.
This condition is equivalent to that the column charge
density between PPF and SS is equal to $E_{\rm f}$ (the Gauss' law).

The net pair flux produced between PPF and a position
$x$ is given  by the multiplication factor, $m(x)$, times the primary
electron flux, $j$. 
The multiplication factor $m(x)$ is obviously a monotonic function
of $x$.
We assume that $\phi$ is also a monotonic function of $x$ between
PPF and SS.
Thus either $x$, $m$ or $\phi$ can be used to designate the
coordinate position.

The charge densities of pairs produced by a small flux element
between $m$ and $m+dm$ at a `position $\phi$' are given by
\begin{equation}
d \bar{\rho}_\pm = \pm j {dm \over \beta_\pm}
=  \pm { j \gamma_\pm \over \sqrt{ \gamma_\pm^2 -1 } } dm,
\eq{drho}
\end{equation}
where the Lorentz factors $\gamma_\pm$ are measured at
the position $\phi$ and are
functions of the injection point denoted by $\phi (m)$, i.e.,
\begin{equation}
\gamma_- = \phi - \phi (m) + \bar{\gamma}
\end{equation}
for electrons and
\begin{equation}
\gamma_+ = - \phi + \phi (m) + \bar{\gamma}
\end{equation}
for positrons; 
$\bar{\gamma}$ is the Lorentz factor
at creation, and assumed to be a universal constant, say 
$\bar{\gamma} \sim 10^2$.

Positrons return back to the star when
\begin{equation} \eq{9}
\phi_{\rm ret} (m) \equiv \phi (m) + \bar{\gamma} -1 < \phi_{\rm s},
\end{equation}
where $\phi_{\rm ret}(m)$ indicates the returning `position',
otherwise positrons have enough energy to go through SS.
Thus there is a separatrics $M_1$ below which
the positron flux ($0 <m<M_1$) returns while beyond which
the positron flux ($M_1 < m < M_2$) passes through SS,
where $M_2$ is the multiplication factor at SS.

The potential at which the flux $M_1$ 
is created is given by the marginal condition of (\ref{9}) 
\begin{equation}
\phi_1 \equiv \phi (M_1)= \phi_{\rm s} - \bar{\gamma} + 1.
\eq{e7}
\end{equation}

If there are returning positrons,
the space charge before PPF becomes 
$-(j/\beta_1)+j_0+jM_1 $, so that positive contribution is added
by the returning flux.
For acceleration of the primary electrons,
the column charge density between the star surface and PPF must 
be negative.
If too many positrons are reflected back, the negative column
density below PPF will be broken down.
If $M_1$ is given and is small enough to assure the negativeness 
of the column charge below PPF, 
the growth of the electric field before PPF is
calculated for an assumed $j$ 
as in the conventional space-charge-limited flow.

Once the primary particles become relativistic, the space charge
before PPF is represented by $-j+j_0+jM_1 $.
If this space charge is negative, then we will have further
acceleration, while if it is positive, then screening.
We may have $-j+j_0+jM_1 >0$, for instance, on field lines 
curving toward the rotation axis.
The positiveness does not guarantee vanishing of $E_\parallel$ at PPF
because PPF may be formed at a place where $E_{\parallel}$ still remains
unscreened.
For this case, the present analysis gives the screening distance after
PPF.
On the other hand, the PPF can be formed in a negative space charge
region where 
\begin{equation} \eq{C0}
-j + j_0 + j M_1 \le 0.
\end{equation}
This takes place 
on field lines with away curvature or with super-GJ current
densities (for any type of field line curvatures). 
For this case, screening relies on pair polarization.


It is notable that the observed X-ray flux from the heated polar cap
suggests that $M_1$ is much less than unity.

In addition to the screening condition (2), we have
a supplementary condition that the second derivative
of $\phi$ must not be positive on SS:
because $d \phi /dx$ is positive at $x=0$ and zero at SS,
if $d^2 \phi /dx^2$ were positive at SS, $d \phi /dx$ would be
negative just below SS, indicating $d \phi / dx$ vanishes
inside SS, 
which contradicts the definition of SS.
Integration of (\ref{drho}) up to $M_2$ yields the pair charge
density at SS, so that
we have positiveness of the effective space charge density at SS as
\begin{equation} \eq{cond3}
\rho_{\rm s} =
- j \int_0^{M_2} {dm \over \beta_-}
+ j \int_{M_1}^{M_2} {dm \over \beta_+}
- (j-j_0) 
\ge 0.
\end{equation}

With the help of (\ref{drho}), the contribution of pairs to
`column charge' in the $\phi$-coordinate
is given by
\begin{eqnarray}
& & \int_0^{\phi_{\rm s}} ( \bar{\rho}_+ + \bar{\rho}_- ) d \phi
\nonumber \\
& = & \ \ \ 
\int_0^{M_1} j dm \int_{\phi (m)}^{\phi_{\rm ret}(m)} {d \phi \over \beta_+}
\nonumber \\
& & \ \ 
+ \int_0^{M_1} j dm \int_0^{\phi_{\rm ret}(m)} {d \phi \over \beta_+}
\nonumber \\
& & \ \ 
- \int_0^{M_1} j dm \int_{\phi (m)}^{\phi_{\rm s}} {d \phi \over \beta_-}
\nonumber \\
& & \ \ 
+\int_{M_1}^{M_2} j dm \int_{\phi (m)}^{\phi_{\rm s}} 
\left[ {1 \over \beta_+}- {1 \over \beta_-} \right] d \phi.
\eq{cdp}
\end{eqnarray}

In the above equation,
the first term comes from the up-going positrons which eventually stops
at $\phi_{\rm ret}$,
the second term is due to the returning positrons.
The third term is for the electrons which are paired with the reflected
positrons. 
The fourth term is for the pairs which pass through the region.

The first term is integrated exactly to yield
\begin{eqnarray}
\left(\begin{array}{c}\mbox{1st term}\\ e^+ \; \rightarrow \end{array} \right)
& = &
\int_0^{M_1} j dm \int_{\phi (m)}^{\phi_{\rm ret} (m)}
{ \gamma_+ \over \sqrt{ \gamma_+^2 -1} } d \phi
\nonumber \\
& = & j M_1 \sqrt{\bar{\gamma}^2 - 1}
\nonumber \\
& \approx &
j M_1 \bar{\gamma} - j M_1 \Delta_1 ,
\end{eqnarray}
where the last expression is obtained if $\bar \gamma \gg1$, 
and $\Delta_1 = (2 \bar{\gamma})^{-1} \ll 1$.
The second terms becomes
\begin{eqnarray}
\left(\begin{array}{c}
\mbox{2nd term} \\ \leftarrow \; e^+
\end{array} \right)
& = & 
j\int^{M_1}_0 jdm \sqrt{(\bar{\gamma}+\phi(m))^2-1},
\nonumber \\
&\approx& 
j \int^{M_1}_0 dm 
  \left[(\bar{\gamma}+\phi(m))-
     \frac{1}{2(\bar{\gamma}+\phi(m))}
  \right] 
\nonumber \\
&=& 
j M_1 \bar{\gamma}+j\int^{M_1}_0\phi(m)dm-
j M_1 \Delta_2 
\eq{2t}
\end{eqnarray}
with
\begin{equation}
\Delta_2 = 
\left\langle \frac{1}{2(\bar{\gamma}+\phi(m))}\right\rangle^{M_1}_0
\ll 1,
\end{equation}
where $\left\langle f(m)\right\rangle_{m_1}^{m_2}$ indicates an
average with respect to the flux $m$, defined as
\[ 
\left\langle f(m)\right\rangle^{m_2}_{m_1} =
\frac{1}{m_2-m_1} \int^{m_2}_{m_1}dm f(m). 
\]

The third term of (\ref{cdp}) is the column charge by electrons 
associated with the returning positrons:
\begin{eqnarray}
\left(\begin{array}{c}
\mbox{3rd term} \\ e^- \; \rightarrow 
\end{array} \right)
& = &
- j\int^{M_1}_0 dm 
  \left[ \sqrt{(\bar{\gamma}+\phi_{\rm s}-\phi(m))^2-1} \right.
\nonumber \\
&   & \ \ \
  \left. - \sqrt{\bar{\gamma}^2-1} \right] 
\nonumber \\
& \approx & 
- j \int_0^{M_1} [ \phi_{\rm s} - \phi (m) ] dm
+ j M_1 \Delta_3
\eq{3t}
\end{eqnarray}
with
\begin{equation}
\Delta_3 =
\left\langle \frac{1}{2\gamma_{\rm -s}} \right\rangle^{M_1}_0 
- \frac{1}{2\bar{\gamma}}
\ll 1,
\end{equation}
where  $\gamma_{\rm -s}(m) \equiv \bar{\gamma}+\phi_{\rm s}-\phi(m)$
is the Lorentz factor of pair electrons at SS.

The fourth term becomes
\begin{eqnarray} 
\left(\begin{array}{c}
\mbox{4th term} \\ e^\pm \; \rightarrow 
\end{array} \right)
&=&  
j\int^{M_2}_{M_1} dm 
\left[ 2\sqrt{\bar{\gamma}^2-1}- \sqrt{\gamma_{\rm -s}^{~2}-1} \right.
\nonumber \\
&  & \ \ \ 
\left. -\sqrt{\gamma_{\rm +s}^{~2}-1} \right],
\eq{4t}
\end{eqnarray}
where, $\gamma_{\rm +s}(m) \equiv \bar{\gamma}-\phi_{\rm s}+\phi(m)$
is the Lorentz factor of the positrons at SS.
If one assumed 
$\bar{\gamma}$,
$\gamma_{- \rm s}$ and
$\gamma_{+ \rm s}$ $\gg 1$,
then the integrand would be
$\bar{\gamma} - \gamma_{- \rm s} + \bar{\gamma} - \gamma_{+ \rm s}$
which would vanish after inserting the definition of 
$\gamma_{- \rm s}$ and
$\gamma_{+ \rm s}$.
This means simply that pairs passing through with relativistic speeds
would not cause any space charge. However,
the positron flux near $M_1$ causes positive net charge because
$\gamma_{+ \rm s} \sim 1$ near $\phi \sim \phi_{\rm s}$,
so that the 4th term cannot be ignored:
we rearrange (\ref{4t}) as
\begin{eqnarray} 
\left(\begin{array}{c}
\mbox{4th term} \\ e^\pm \; \rightarrow 
\end{array} \right)
& = &
2 j \int_{M_1}^{M_2} [ \sqrt{\bar{\gamma}^2 -1} - \bar{\gamma} ] dm
\nonumber \\
&   &
- j \int_{M_1}^{M_2} [ \sqrt{\gamma_{+ \rm s}^2 -1}- \gamma_{+ \rm s} ] dm
\nonumber \\
&   &
- j \int_{M_1}^{M_2} [ \sqrt{\gamma_{- \rm s}^2 -1}- \gamma_{- \rm s} ] dm,
\eq{term4}
\end{eqnarray}
which can be integrated exactly (see Appendix)
if the pair creation rate is      constant
with respect to the `$\phi$-coordinate'.
In this approximation, we have
\begin{equation}
m  =  M_1 + \alpha_\phi \left[ \phi (m) - \phi_1  \right]
\end{equation}
with
\begin{equation}
\alpha_\phi  =  
{M_2 - M_1 \over \phi_{\rm s} - \phi_1 }
= 
{M_2 - M_1 \over \bar{\gamma} - 1 }
= \mbox{const.},
\eq{alph}
\end{equation}
where $\alpha_\phi$ is the pair multiplication factor 
created in a unit potential drop $mc^2/e$,
in other words, $\alpha_\phi$ is the pair multiplication factor produced 
in the `braking distance' $\Delta s =mc^2/e |E_\parallel |$.
After integration we have 
\begin{equation} \eq{22}
\left(\begin{array}{c}
\mbox{4th term} \\ e^\pm \; \rightarrow 
\end{array} \right)
=
j \alpha_\phi \zeta ,
\end{equation}
where $\zeta$ is a constant of order of unity;
$\zeta = 1.744$ for $\bar{\gamma}=100$
(see Appendix for derivation).

Near SS, $d \phi / dx \rightarrow 0$, $\alpha_\phi=dm/d\phi$
is not constant if the pair creation is spatially uniform.
In the next section,  we perform a numerical calculation for
a spatially uniform pair creation and find that the above evaluation is quite
accurate.
The analytical evaluation by using a constant rate 
$\alpha_\phi=dm/d\phi$ will be appropriate for more or less uniform
pair creation.

Inserting the above evaluation of all the four terms into (\ref{Poisson2})
and rearranging relevant terms,     we have
\begin{equation}
{1 \over 2} E_{\rm f}^2 =
2jM_1 - 2jM_1 \langle \Delta \phi \rangle
+ j \alpha_\phi \zeta
- (j-j_0-jM_1) \phi_{\rm s}
\eq{Poisson3}
\end{equation}
where 
\begin{eqnarray}
\langle \Delta \phi \rangle
& = &
{1 \over M_1 } \int_0^{M_1} [ \phi_{\rm s} - \phi_{\rm ret} (m) ] dm
\nonumber \\
& = &
{1 \over M_1 } \int_0^{M_1} [ \phi_1 - \phi (m) ] dm
\end{eqnarray}
is the potential drop (flux weighted) in the region where returning lasts,
in other words, 
the potential difference between
where returning start and where screening completes.

The pair effect is made up of three components corresponding 
to the first three terms in (\ref{Poisson3}). 
The first effect is a positive charge produced by non-relativistic 
part of the  returning positrons.
The second effect is negative because it is due to
electrons left behind the returning positrons. 
This effect by returning is fatal: the more positrons return,
the more electrons are left behind to create negative space charge
in the downstream (outer) region,
whereas the region needs positive charge for screening.
The last effect is caused by
polarization of the  pairs 
which pass through SS (3rd term).

If the region of returning is large,
in other words,
if $\langle \Delta \phi \rangle \gg 1 $,
the pair screening becomes obviously difficult to complete because
the negative space charge produced by pair electrons left behind
in upstream region becomes large to prevent screening.
If uniform pair creation is assumed,
$\langle \Delta \phi \rangle = (1/2) \phi_1$, 
and $M_1 = \alpha_\phi \phi_1$.
In this case, the condition (\ref{Poisson3}) becomes
\begin{equation} \eq{27}
{1 \over 2} E_{\rm f}^2 =
j \alpha_\phi [ \phi_1 (2-\phi_1) + \zeta ]
- (j-j_0-jM_1) \phi_{\rm s}.
\end{equation}
The first term of the right hand side of (\ref{27}) corresponds to the 
pair polarization.
The second term is the `background charge', the sign of which
depends on the global parameters $j$ and $j_0$, and
where PPF is formed.
Let us restrict ourself to the case for which the global effects is
not helpful but pairs cause screening, i.e., 
the last term makes no positive contribution.
In order for the right hand side of (\ref{27}) to be positive,
the necessary condition is that the pair polarization term
is positive, and then we have
\begin{equation}
2 \langle \Delta \phi \rangle = \phi_1 < 1 + \sqrt{1 + \zeta}
\approx 2.6,
\end{equation}
which means that returning of positrons lasts only
a few braking distance.

Here we invoke the condition (\ref{cond3}) that is for charge             
density at SS to be positive:
\begin{eqnarray}
\rho_{\rm s} &=& -j\int^{M_2}_0 \frac{dm}{\beta_-}
+j\int^{M_2}_{M_1}
\frac{dm}{\beta_+}-(j-j_0) \nonumber \\
&=& j\alpha_\phi\left( 2\sqrt{\bar{\gamma}^2-1}
-\sqrt{(2\bar{\gamma}+\phi_1-1)^2-1} \right)
-(j-j_0) \nonumber \\
&\approx& (1-\phi_1)\alpha_\phi j -(j-j_0) \nonumber \\
& = & j \alpha_\phi (1-2 \phi_1) -(j-j_0-M_1j) > 0.
\eq{e26}
\end{eqnarray}
As was mentioned, returning positrons leave negative space
charge at SS, the positiveness of $\rho_{\rm s}$ sets an strong constraint:
from (\ref{e26}), again noting no positive contribution from the last
term, we have
$(1-2\phi_1)>0$, i.e.,
\begin{equation}
0 < \phi_1 < {1 \over 2}.
\eq{cnd2}
\end{equation}

The condition (24)
is then rewritten as
\begin{equation}
{1 \over 2} E_{\rm f}^2 =
j \alpha_\phi \zeta^\prime - (j-j_0-jM_1 )\phi_{\rm s}
\eq{efinal}
\end{equation}
where $\zeta^\prime$ is  between $\zeta = 1.744 $ (when $\phi_1=0$)
and $\zeta + 3/4 = 2.494$ (when $\phi_1=1/2$).
Eventually the necessary condition for pair polarization to screen out
$E_{\rm f}$ becomes
\begin{equation} \eq{32}
\alpha_\phi \ge {E_{\rm f}^2 \over 2 j \zeta^\prime }.
\end{equation}
Recall that $\alpha_\phi$ is the pair multiplication factor produced
{\it within the braking distance} $\Delta s = mc^2 /e | E_{\parallel}|$.
In dimensional form,
the required pair multiplication factor in $\Delta s$ is given by
\begin{equation} \eq{30}
\Delta M_{\rm screen} \ge
{ E_{\parallel}^2 \over 8 \pi mc^2 n_0 \zeta^\prime j }.
\end{equation}
For polar cap models,
typical values of the right hand side of (\ref{30}) are as large as
$10^3$. 
For instance, given a primary acceleration
of $\gamma \sim 10^6$ in a distance of $L \sim 10^4$cm,
we may have $|E_{\parallel}| =1.7  \times 10^5$ in esu or                       $E_{\parallel}^2/8 \pi= $ 
$1.2 \times 10^9 (\gamma_6 / L_4)^2$ erg cm$^{-3}$ at PPF, and
\begin{equation}
\Delta M_{\rm screen} \ge
2.0 \times 10^3 ({\gamma_6 \over L_4})^2
({P_{0.1} \over B_{12}})
{ 1 \over j \zeta^\prime },
\end{equation}
where $\gamma_6=\gamma / 10^6$ , and 
$L_4$, $P_{0.1}$ and $B_{12}$  are
in units of $ 10^4$ cm, 0.1 sec and $10^{12}$G, respectively.
This multiplication factor should be achieved within a
distance of $\Delta s=mc^2 / e|E_{\parallel}|=                                  \gamma^{-1} L \approx 10^{-2} $ cm
for screening. 
It seems quite difficult that
the pair multiplication factor of order of $10^3$ 
in this sort of small distance.
If the pair production takes place over a dimension of 
say $10^4-10^5$ cm,
then the required multiplication factor in the whole region
is $10^9-10^{10}$ and is much higher
than the value predicted by conventional magnetic pair production.
If, on the other hand, we assume $L \approx 10^6$ cm,
then $\Delta M_{\rm screen} \approx 0.2$, 
$\Delta s = 1$ cm and
the total multiplication factor becomes 2000 to 20000, which may be
allowed.

In the case where $j<j_0$, the screening completes with the help of
the `background' positive space charge, i.e.,
$-(j-j_0-jM_1) \phi_{\rm s}$ in (\ref{27}).
However, the distance to the screening region is enlarged by
the negative contribution due to the pair electrons left behind.
For a given PPF, i.e., given $j$, $j_0$ and the pair creation rate
$\alpha_\phi$, we can calculate the screening distance
in potential scale $\phi_{\rm s}$ and return flux $M_1$ in use of
(\ref{27}), (\ref{e7}) and $M_1=\alpha_\phi \phi_1$:
\begin{equation}
\phi_{\rm s} =
{(1/2)E_{\rm f}^2 + j \alpha_\phi ( \bar{\gamma}^2 - 1 - \zeta )
\over
j_0 - j [1 - \alpha_\phi (\bar{\gamma} + 1) ] }.
\end{equation}
It can be seen that the screening is possible if
$j < j_0 /[1-\alpha_\phi (\bar{\gamma} +1)] \equiv j_{\rm cr}$.
For the conventional polar cap model, $j_{\rm cr} \approx j_0$
because $\alpha_\phi \bar{\gamma}$ is much less than unity.
If $j \rightarrow j_{\rm cr}$, $M_1$ becomes large so that
we have to treat the whole acceleration region including
the cathode region, which will be discussed in a subsequent paper.

\section{Numerical Results}

In the previous section, we have discussed the screening
condition under the assumption that
the pair creation rate is constant with respect to the electrostatic potential.
In this section, we present numerical solutions 
when the pair creation rate is spatially uniform,
$dm/dx=$ constant.

First we check the code by calculating the same
case as in the previous section.
We present the case of $\bar{\gamma}=100$, $j=2j_0$,
$\alpha_\phi=1$ and $M_1=0.5$,
and hence $\phi_1=0.5$.
The results are shown in Figures~3 and 4, where
$\phi$ and $|E|$ as functions of $x$ are shown.
We find that the screening of the electric field
occurs in the region where non-relativistic
positrons appear, and the distance scale of this region
is roughly the braking scale of $mc^2/e|E_{\parallel}|$.
The whole dimension between PPF and SS is ~$\bar{\gamma}$
in potential scale. 
Consequently 99\% of the pairs
produced in this region go through and only 1\%
of positrons return back to the star.
In this case, the numerical calculation gives $E_{\rm f}=3.159$, which
perfectly agrees with the numerical estimate based on (\ref{32}),

\begin{figure}
\begin{center}
\mbox{\epsfysize=6cm \epsfbox{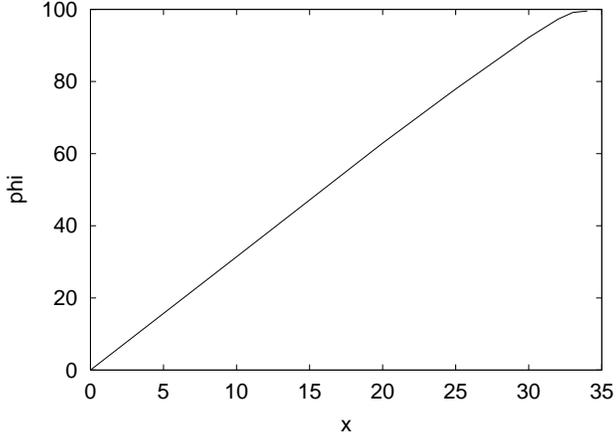}}
\end{center}
\caption{%
Electrostatic potential as a function of the distance 
from the pair production front.
}
\end{figure}

\begin{figure}
\begin{center}
\mbox{\epsfysize=6cm \epsfbox{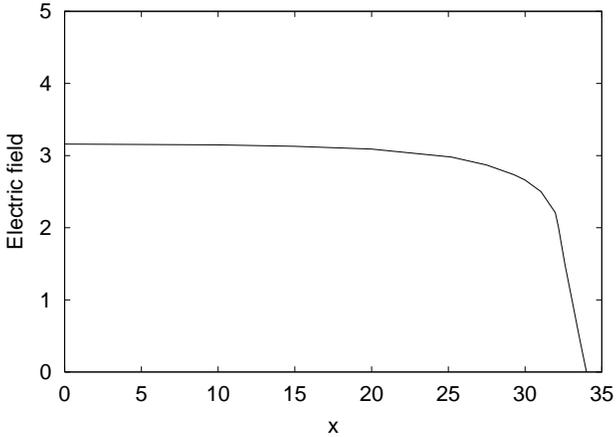}}
\end{center}
\caption{
Electric field strength as a function of the distance 
from the pair production front.
}
\end{figure}

Next, we examine the case where the pair creation rate
is spatially constant, i.e., $m(x)=\alpha_x x$ with a
constant $\alpha_x$. For $\bar{\gamma}=100$, $j=2j_0$,
and $M_1=0.5$, we solve (\ref{Poisson1}) numerically.
When $\phi_1=0.5$, we obtain the electric
field strength at PPF, $E_{\rm f}=3.160$, which is almost the same value as
the case of the constant $\alpha_\phi$. 
If one plots as in the same diagram as Figures~3 and 4, it is hard to
find the difference on the plots.

The small difference found in $E_{\rm f}$ between the
two cases arises from the
behavior of the pair creation rate near $\phi=\phi_{\rm s}$.
When we assume $m=\alpha_\phi \phi$, the spatial pair creation
rate becomes slightly smaller as the electric field becomes weaker
near SS. With this numerical solution, we find that (\ref{32}) is
useful even for the case of the spatially uniform pair creation.

\begin{figure}
\mbox{\epsfysize=6cm \epsfbox{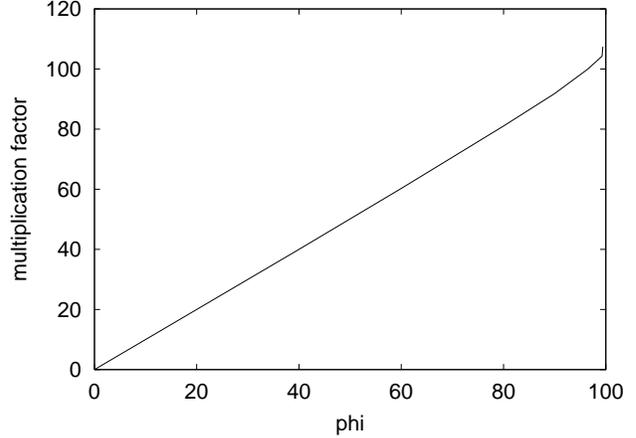}}
\caption{%
Multiplication factor as a function of $\phi$ when  
${dm}/{dx}$ is constant. 
}
\end{figure}

\section{Concluding remarks}

In this paper, we considered the electric field screening by
polarization of electron-positron pairs which are
created beyond the pair production front as is
generally supposed for the polar cap models. 
We allow the pair positrons to return back to the star as far as
the condition for the acceleration is satisfied.
The one-dimensional Poisson equation is solved together with particle
motion for a steady state acceleration region.
We obtain the formula (\ref{27}) which give the electric field
strength that can be screened for a given current density $j$,
a given field geometry $j_0$ and a given pair creation rate
$\alpha_\phi$.
It is found that 
(1) pair polarization has little contribution for positive space
charge, and
(2) returning of positrons makes the screening difficult
seriously because the pair electrons left behind the returning positrons
produce negative space charge in the screening region where the positive
space charge is required.
As a result, the thickness of the screening is restricted to be as small as
the braking distance $\Delta s \approx mc^2 / e | E_{\parallel} |$ 
for which positrons become non-relativistic.
We confirmed the previous result of Paper I that the electric field
in the case where geometrical screening is not possible, i.e.,
$-j+j_0+jM_1 <0$ at the pair production front,
we have, from (\ref{30}), 
\begin{equation}
{E_\parallel^2 \over 8 \pi } <
mc^2 \left( \Omega B \over 2 \pi c e \right) \zeta^\prime j
\Delta M_{\rm screen}
\end{equation}
where $\Delta M_{\rm screen}$ is the pair multiplication factor
{\em within} $\Delta s$.
If the primary current density is of order of the Goldreich-Julian (GJ) value,
the required pair multiplication factor per one primary electron is 
enormously large and cannot be realized in the conventional pair
creation models.
A previous belief that pair creation
with a pair density higher than the GJ density immediately
screens out the electric field is unjustified.

Some mechanism to salvage this difficulty should be found.
As already mentioned,
if the last term of (\ref{27}), 
$-(j-j_0 - j M_1) \phi_{\rm s}$, becomes positive
screening is possible as far as $j<j_{\rm cr} \approx j_0$.
Since $j_0 = B_{\rm z}/B <1$, $j$ must be less than unity.
It must be noted that the positive space charge exists before
the pair production front in this case, 
so that screening takes place before the pair production front and
$E_{\parallel}$ is reduced. Even a reversal of $E_\parallel$
is possible for a small $j$ (Shibata 1995).
In any case, if geometrical screening works,
a finite potential drop along field lines can complete.
But the current density is restricted in a certain range.

The condition (\ref{30}) implies that 
if the current density is much higher than the GJ value,
the required multiplication factor can be reduced.
Although the total current circulating in the magnetosphere
should be of order of so-called GJ current, 
$I_{\rm GJ} =\mu \Omega^2 / c$, where $\mu$ is the magnetic
moment of the star, current densities can be in principle focused:
the GJ current density is just a value if 
$I_{\rm GJ}$ is
more or less uniformly distributed over the whole polar cap area,
and therefore if the current distributed, say, in an annular region 
with fine structures such as seen in aurora, the focused current density
will be much higher than the GJ current density.
Thus, if $j$ is much larger than $j_0$, and the required multiplication factor
will be reduced.
(If the current is super-GJ, toward curvature will not helpful for
screening because $-(j-j_0 - j M_1)$ is negative.)
In any case,
physics in focused field-aligned current,
e.g., effects of inhomogeneity across the magnetic field lines,
two stream instability in a high current density with unscreened 
electric field, 
has yet been properly studied. 

Since $\bar{\gamma}$ is an order of 100, $\phi_s$ turns out to be
an order of 100 using (\ref{e7}) and (\ref{cnd2}). 
From (\ref{alph}) we may
safely approximate as $\alpha_\phi = M_2 /100$. 
From (\ref{32}),
$E_{\rm f} \le \sqrt{M_2}/5$, taking $\zeta^\prime =2$ and $j=1$. 
As an example,
let us assume that total pair production occurs with a multiplication factor
$10^4$ over the distance $x_{\rm p}=10^4$.
Thus, $M_2 = 10^4 (x_{\rm s}/x_{\rm p})=x_{\rm s}$. 
Noting that $E_{\rm f} \approx \phi_{\rm s} /x_{\rm s}=100/x_{\rm s}$,
we obtain $x_{\rm s} > 60$ and $E_{\rm f } < 1.5$.
Since the unit length $c/\omega_{\rm p}$ is typically 1 cm,
this electric field corresponds to a voltage $mc^2/e$ per cm.
Assuming that primary electrons are accelerated up to $\gamma> 10^6$
with this weak field $E_{\rm f} \sim 1$, the required length
between the neutron star surface and the pair production front is about $10^6$ cm
in hight.
Thus pair screening is possible
if the pair production front is located high
above the surface and if the field-aligned electric field is not so strong.
However, the previous models predict a stronger electric field,
so we would need an additional ingredient to keep a small field
as long as a stellar radius.
In any case, a weak electric field kept in a long distance
is one possible solution to have a self-consistent polar cap model.

Another possible way out is to include frictional forces
between various components of charged currents.
Since friction on positrons pulls them outwards
along with electrons, returning fraction of positrons will be reduced,
which will make screening easier.
Although there are some ideas for physical processes of friction
(two stream instability, production of positroniums and others),
it is not clear whether the frictional force can be strong enough
to lift positrons and screen the electric field.
Study on the frictional interaction under {\it unscreened} 
electric field is strongly demanded.

\subsection*{Acknowledgments}

We thank K. Masai, D. Melrose and Q. Luo for useful comments and discussions.
This work is supported in part by Grant-in-Aid for Scientific
Research from the Ministry of Education, Science, and Culture of
Japan Nos. 12640229 and 13440061.

\begin{center}{\bf References} \end{center}
\begin{verse}
\footnotesize

Arons, J., Scharlemann, E.T., 1979, {\it Ap. J}, {\bf 231}, 854

Fawley, W.M., Arons, J., Scharlemann, E.T., 1977, {\it Ap. J}, {\bf 217}, 227 

Hirotani, S., Shibata, S., 2001, {\it Ap. J.}, {\bf 558}, 216

Mestel L., 1981, IAU Symp.~95: Pulsars: 13 Years of Research on Neutron Stars, 95, 9. 

Miyazaki, J.\& Takahara, F., 1997, {\it MNRAS}, {\bf 290}, 49

Muslimov, G.A. and Tsygan, A.I., 1992, {\it MNRAS}, {\bf 225}, 61

Scharlemann, E.T., Arons, J., Fawley, W.M.,  1978, {\it Ap. J}, {\bf 222}, 297

Shibata, S., 1991, {\it Ap. J},  {\bf 378}, 239

Shibata, S., 1995, {\it MNRAS}, {\bf 276}, 537

Shibata, S., 1997, {\it MNRAS}, {\bf 287}, 262

Shibata, S., Miyazaki, J. \& Takahara, F., 1998, {\it MNRAS}, {\bf 295}, L53

\end{verse}

\appendix

\section*{Appendix A: Space charge by pairs go through the screening region}

The space charge by pairs that go through the screening region is
given by (\ref{term4}).
The second and third terms can be integrated analytically if
the pair flux is proportional to the potential for which
\begin{equation}
dm = \alpha_\phi d \phi (m)
\end{equation}
where the pair creation rate in unit potential drop $\alpha_\phi$
is assumed to be constant.

The second term of (\ref{term4}) becomes
\begin{eqnarray}
&  &
j \int_{M_1}^{M_2} dm 
\left( \gamma_{+\rm s} - \sqrt{\gamma_{+ \rm s}^2 -1 } \right)
\nonumber \\
& = &
j \int_{\phi_1}^{\phi_{\rm s}}
\left( \gamma_{+\rm s} - \sqrt{\gamma_{+ \rm s}^2 -1 } \right)
\alpha_{\phi} d \phi (m) 
\nonumber \\
& = &
{1 \over 2} j \alpha_\phi
\left[
\bar{\gamma}^2 - \bar{\gamma} \sqrt{ \bar{\gamma}^2 -1 }
+ \ln ( \bar{\gamma} + \sqrt{\bar{\gamma}^2 - 1 }) - 1 
\right].
\end{eqnarray}
In the same way, the third term of (\ref{term4}) becomes
\begin{eqnarray}
&   & 
j \int_{M_1}^{M_2} dm 
\left( \gamma_{- \rm s} - \sqrt{\gamma_{- \rm s}^2 -1 } \right)
\nonumber \\
& = &
j \int_{\phi_1}^{\phi_{\rm s}}
\left( \gamma_{- \rm s} - \sqrt{\gamma_{- \rm s}^2 -1 } \right)
\alpha_{\phi} d \phi (m) 
\nonumber \\
& = &
{1 \over 2} j \alpha_\phi
\left[
\gamma_{- \rm s}^2 - \gamma_{- \rm s} \sqrt{ \gamma_{- \rm s}^2 -1 }
\right.
\nonumber \\
& & \ \ 
\left.
+ \ln ( \gamma_{- \rm s} + \sqrt{\gamma_{- \rm s}^2 - 1 }) - 1 
\right]_{\bar{\gamma}}^{2 \bar{\gamma} -1}.
\end{eqnarray}

Thus we arrive at (\ref{22}) with $\zeta = 1.744$ if 
$\bar{\gamma} = 100$ is substituted.

\bsp

\end{document}